\documentclass[conference]{IEEEtran}
\usepackage{graphicx,times,amsmath,booktabs,algorithm,algorithmic,amsmath}
\usepackage{multicol}
\usepackage{amsmath,amssymb}
\usepackage{subfigure}
\usepackage{stfloats}
\usepackage{epstopdf}

\usepackage{balance}

\newtheorem{theorem}{Theorem}
\newtheorem{lemma}{Lemma}

\begin{document}

	\title{Coverage Analysis of Downlink Transmission in Multi-Connectivity Cellular V2X Networks }
	

	\author{ \IEEEauthorblockN{Luofang Jiao\IEEEauthorrefmark{1}, Tianqi~Zhang\IEEEauthorrefmark{1}, Jiwei Zhao\IEEEauthorrefmark{1}, Yunting~Xu\IEEEauthorrefmark{1}, and Haibo Zhou\IEEEauthorrefmark{1}}
		\IEEEauthorblockA{\IEEEauthorrefmark{1}School of Electronic Science and Engineering, Nanjing University, Nanjing, China, 210023 }
		Email: {luofang\_jiao@foxmail.com, tianqizhang@smail.nju.edu.cn, jw\_zhao@smail.nju.edu.cn,}\\ {yuntingxu@smail.nju.edu.cn, haibozhou@nju.edu.cn} \\	
	}

	\maketitle
	\begin{abstract}
		With the increasing of connected vehicles in the fifth-generation mobile communication networks (5G) and beyond 5G (B5G), ensuring the reliable and high-speed cellular vehicle-to-everything (C-V2X) communication has posed significant challenges due to the high mobility of vehicles. For improving the network performance and reliability, multi-connectivity technology has emerged as a crucial transmission mode for C-V2X in the 5G era. To this end, this paper proposes a framework for analyzing the performance of multi-connectivity in C-V2X downlink transmission, with a focus on the performance indicators of joint distance distribution and coverage probability. Specifically, we first derive the joint distance distribution of multi-connectivity. By leveraging the tools of stochastic geometry, we then obtain the analytical expressions of coverage probability based on the previous results for general multi-connectivity cases in C-V2X. Subsequently, we evaluate the effect of path loss exponent and downlink base station density on coverage probability based on the proposed analytical framework. Finally, extensive Monte Carlo simulations are conducted to validate the effectiveness of the proposed analytical framework and the simulation results reveal that multi-connectivity technology can significantly enhance the coverage probability in C-V2X.

	\end{abstract}
	\IEEEpeerreviewmaketitle

	
	\section{Introduction}


With the evolutionary development of vehicular networks,  cellular vehicle-to-everything (C-V2X) is emerging as a key technology for improving the efficiency and safety of vehicular traffic and enabling various applications that enhance the driving experience and environmental sustainability in intelligent transportation systems (ITS) \cite{wolf2018reliable}. Through leveraging the existing cellular network infrastructure and spectrum, C-V2X is capable of providing effective communication among vehicles (V2V), as well as between vehicles and other network entities (V2I) such as base stations (BSs), roadside units (RSUs), and cloud servers \cite{chen2017vehicle}. However, C-V2X also faces significant challenges due to the high-speed mobility and dynamic topology of vehicles, which may cause rapid fluctuations in the quality of wireless links, frequent handovers, and increased signaling overhead \cite{chen2020vision}. These challenges may result in worse communication performance and influence the quality of experience (QoE) and quality of service (QoS) of ITS applications, such as collision avoidance, traffic management, cooperative driving, platooning, and autonomous driving \cite{jiao2022spectral}. To fully exploit the potential of C-V2X, reliable and high-performance wireless communication systems are extremely essential.

In recent years, multi-connectivity technology has attracted tremendous academic attention and has been widely considered as a promising mechanism to improve communication reliability, reduce latency and boost overall network performance \cite{zhao2023fully, lu2022personalized}.
	Kousaridas \textit{et al.} suggested a 5G  Radio Access Networks (RAN) based approach to enhance the multi-connectivity abilities and expected QoS advantages in \cite{kousaridas2019multi}. 
	By exploiting the diversity of wireless channels and the available radio resources in the cellular network, multi-connectivity can improve the overall system throughput and reliability by combining or switching among different connections based on their quality and availability. It can also reduce the handover latency and signaling overhead by maintaining seamless connectivity during mobility and avoiding frequent connection re-establishment \cite{suer2019multi}.
	
	In the context of multi-connectivity in C-V2X,
	Lu \textit{et al.} \cite{lu2022personalized} introduced a novel approach for reducing the duplication rate in the downlink base stations (DBSs) of the fully-decoupled networks in C-V2X.
	Rabitsch \textit{et al.} \cite{rabitsch2020utilizing} investigated the multi-access algorithms to meet the stringent requirements for communication availability and latency in V2I networks.
	Moreover, Wu \textit{et al.} \cite{wu2020performance} proposed a multi-connectivity scheme for uplink in C-V2X systems. They obtained the exact expressions of the outage probability of uplink by using the tools of stochastic geometry.
	Consequently, multi-connectivity is essential for improving C-V2X performance and supporting various kinds of vehicular applications, such as collision avoidance, traffic management, and autonomous driving.
	However, most of the existing works on multi-connectivity C-V2X communication mainly focused on uplink and optimization. 
	The performance of multi-connectivity C-V2X communication in downlink transmission has not been sufficiently researched on how to analyze it.

	
	To this end, we present a feasible analytical framework for downlink transmission in multi-connectivity C-V2X networks. By modeling the vehicles and DBSs as 1-D Possion point processes (PPPs), stochastic geometry is used to obtain the important performance indicators, such as joint distance distribution and coverage probability. The detailed contributions of this paper are summarized below:
	\begin{itemize}
		\item We present a novel multi-connectivity performance analysis framework in C-V2X, which enables the evaluation of network performance in the 5G era. 
		
		
		\item We derive the precise expressions of joint distance distribution and coverage probability for general multi-connectivity cases in C-V2X. 
		
		\item We conduct comprehensive Monte Carlo simulations to confirm the effectiveness of the presented multi-connectivity performance analysis framework, which show that multi-connectivity technology can significantly improve network performance in C-V2X.

	\end{itemize}
	
	The subsequent sections of this paper are structured as follows. Section \uppercase\expandafter{\romannumeral2} presents the proposed framework for analyzing multi-connectivity performance.
	Section \uppercase\expandafter{\romannumeral3} conducts a performance analysis of the system, including the joint distance distribution and coverage probability.
	In Section \uppercase\expandafter{\romannumeral4}, the simulation setup and results obtained from extensive Monte Carlo simulations are presented, providing verification of the proposed framework and evaluation of the system performance.
	Section \uppercase\expandafter{\romannumeral5} presents the concluding remarks of this paper.
	
	\section{ANALYSIS SCHEME}

	In this section, we introduce a simplified one-dimensional (1-D) system model to study downlink multi-connectivity in the C-V2X scenario.
	In order to implement the spectrum allocation, we leverage a coordination scheme named Single Frequency Networks (SFN) \cite{simsek2019multiconnectivity}, which requires BSs to coordinately create signals and strictly synchronize their timing.
   SFN enables the transmission of incoherent joint signals on the same radio resources in frequency and time.
	Our focus in this paper is on intra-frequency multi-connectivity, which involves the simultaneous transmission of multiple DBSs operating at the same carrier frequency to the same vehicle. This is an important issue to address in the C-V2X scenario, where high data rates and reliable communication are required for safety-critical applications.
	Therefore, we aim to investigate the performance of intra-frequency multi-connectivity in C-V2X and the specific channel model, association policy, interference model, and performance metrics utilized in this paper are elaborated in the following.
	
	
	
	\subsection{Modeling of C-V2X Network}
	
\begin{figure}[t]
		\centering
		
		\centerline{\includegraphics[width=1.0\hsize]{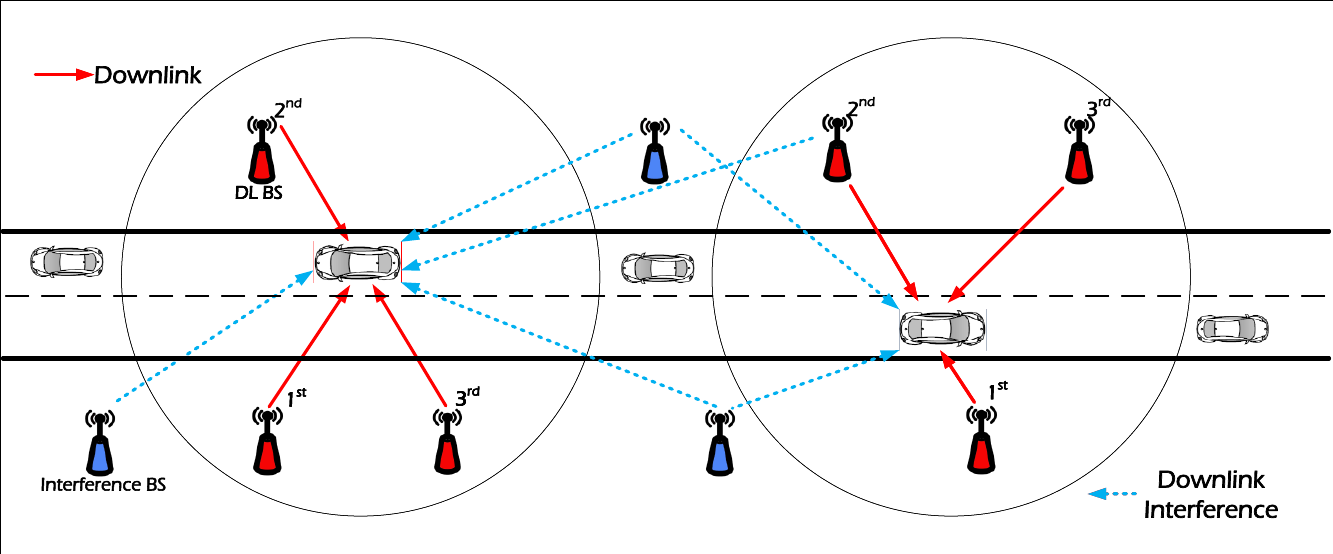}}
		\caption{An example of a practical 1-D scenario for downlink transmission in multi-connectivity C-V2X is illustrated. In this scenario, the target vehicle receives messages from the three closest DBSs, while transmissions from DBSs located beyond the collaboration distance can lead to interference to the target vehicle.}
		\label{1-d system-model}
	\end{figure}

	
	As shown in Fig. \ref{1-d system-model}, an example of downlink multi-connectivity scenarios in C-V2X is illustrated, where vehicles are randomly distributed on an urban freeway segment, and DBSs are densely distributed along the road. To simplify the analysis, we make the assumption that both the DBSs and vehicles utilize a single antenna.
	
	For the tractability of the downlink performance analysis, we consider a 1-D scenario on a road including vehicles, DBSs, and interference DBSs as shown in Fig. \ref{1-d system-model}, as in \cite{wu2020performance}.  We use the 1-D homogeneous Poisson point processes (PPP) $ {\varphi _V}, \varphi _D $ with density $ {\lambda _v}, \lambda _d$ to denote the locations of vehicles, DBSs on the road, respectively. And $ {\varphi _V}, \varphi _D $ can be expressed as
	\begin{align}
	\varphi_j\mathop {=}\limits^{\bigtriangleup}\left \{ x_{i,j}\in R^2: i\in \mathbb{N}_+ \right \},j=\left \{ V, D \right \}.\nonumber 
	\end{align}
	All of the vehicles and DBSs are distributed along a road with length $ l $. As per Slivnyak's theorem, the distribution of point processes remains unchanged even after adding a node at the origin \cite{chiu2013stochastic}, and in order not to lose generality and eliminate segmentation due to boundary effects, we place the typical vehicle at the origin $ v_o=(0, 0) $, i.e. which represents the center of the road.  
	
	In relation to the formation of virtual cells, we assume that each vehicle is connected to the $ n $ nearest DBSs on a Euclidean plane. The 1-D distance between the typical vehicle $ v_o $ with the $ i $-th ($ i\le n $) DBS is $ x_i $.
	We adopt a common power-law pathloss and Rayleigh fading model with a decay rate of $ \left | x \right |^{-\alpha_d } $, where $ x $ denotes the distance between the DBS and the typical vehicle. The downlink pathloss exponent parameter is denoted as $ \alpha_d $ $(\alpha_d>2)$. $ g_d $ is used to denote the power gain of Rayleigh fading and it is modeled by an exponential distribution with a mean of $1/\mu$, where $ \mu $ represents the mean power gain of the fading channel. Therefore, we have $ g_d \sim exp\left( \mu \right) $. The distribution function of $ g_d $ is 
	\begin{align}
	f\left ( x\right ) =\mu  e^{-\mu x}.
	\end{align} 
{Furthermore, we use random variable $ {\chi _{d}} $ to model the effects of shadowing between the DBS and the typical vehicle in the downlink, and $ {\chi _{d}} $  follows a log-normal distribution given by $ 10log_{10}\chi_{d}\sim \aleph \left ( \omega _{d},\delta _{d}^{2} \right ) $, where $\omega_d$ represents the mean of the logarithm of $\chi_d$ (i.e., the geometric mean of $\chi_d$), while $\delta_d^2$ represents the variance of the logarithm of $\chi_d$ \cite{abdulqader2015performance}.}
	Hence, the received signal power of the typical vehicle from $ i $-th DBS in downlink is 
	\begin{equation}
	P_{r,v}(x_i) = {{P_d}{g_{d}}{\chi _{d}}{{\left\| x_i \right\|}^{ - {\alpha _d}}},} ~{x_i \in  {\varphi_D}},
	\end{equation}
	where $ P_d $ is the transmitting power of DBS and we assume that all DBSs have same power $ P_d $.

	\subsection{Association policy} 
	
		The typical vehicle is assumed to be connected to $ n $ DBSs by measuring all the received power from the nearby DBSs, finding the DBSs with max received power (MRP) in turn \cite{wu2020performance}.
	Since the received power $ P_{r,v} $ is not exponentially distributed for the modeling of the shadow fading \cite{jiao2022spectral}, the lemma of the random displacement theorem is considered to solve this issue. 
	Thus, $ P_{r,v}(x_i)={P_d}{g_{d}}{\chi _{d}}{{\left\| x_i \right\|}^{ - {\alpha _d}}}$ can be transformed to $ P_{r,v}(y_i)={P_d}{g_{d}}{{\left\| y_i \right\|}^{ - {\alpha_d }}}  $, where $ y_i={\chi_{d}^{-\frac{1}{\alpha_d }}}x_i $. 
	The 1-D PPP transformed converges to a 1-D homogeneous PPP and the intensity $ \lambda $ is transformed to $ \mathbb{E}\left [ {\chi_{}^{-\frac{1}{\alpha }}}  \right ]\lambda $, and the intensity of the 2-D PPP is $ \mathbb{E}\left [ {\chi_{}^{-\frac{2}{\alpha }}}  \right ]\lambda $ after executing the procedure of random displacement \cite{chetlur2018coverage}. Specifically, the $ \mathbb{E}\left [ {\chi_{}^{-\frac{1}{\alpha }}}  \right ]\lambda $ can be calculated as 
	\begin{align}
	\mathbb{E}\left[ {{\chi ^{ - \frac{1}{\alpha }}}} \right]\lambda  = \exp \left( {\frac{{\omega \ln 10}}{{10\alpha }} + \frac{1}{2}{{\left( {\frac{{\sigma \ln 10}}{{10\alpha }}} \right)}^2}} \right)\lambda,
	\end{align}
	where $ \mathbb{E}\left ( \cdot  \right )  $ is the expectation function.
	Then we use 1-D PPP $ \varphi _{_D}^t $ to denote the transformed set of DBS and $ \lambda_{D} = \mathbb{E}\left [ {\chi_{}^{-\frac{1}{\alpha_d }}}  \right ]\lambda_d  $ denotes the transformed DBS intensity. 
Thus the candidate serving DBSs is changed to the  $ n $ nearest DBSs $ \in  \varphi _D^t $ in turn, and this can be expressed as
	\begin{align}
	P_{r,v}(x_i) = \mathop {\arg \max }\limits_{i \in {\varphi _D^t\backslash\varphi _c  }} {P_d}{g_d}x_i^{ - {\alpha _d}},i>m, 
	\end{align}
	where we use $ \varphi_c=\{x_1,x_2, \dots, x_{m}\} $ to denote the connected collaborative DBSs set and $ x_{i},{ i \in \{m+1, m+2, \cdots\}}$ denotes the distance between the $ i $-th nearest DBS $\in \varphi _D^t \backslash\varphi _c$ and the typical vehicle. This means that to expand the set of collaborative DBSs set $ \varphi _c $, we need to find the nearest DBS among DBSs outside the connected collaborative DBSs set.  
	\vspace{-0.1cm}
	\subsection{Interference}
	
	In the collaboration DBSs set $ \varphi _c $, all DBSs will transmit the control and data signals simultaneously on the same subband.
	Since the signal components of the DBSs are within the cyclic prefix, the resulting multi-connectivity signal to interference plus noise ratio (SINR) experienced by the typical vehicle $v_o$ in downlink is defined as follows:

	\begin{align} \label{SINR}
	S\!I\!N\!{R_D} = \frac{{\sum\limits_{i \in \varphi _c} {{P_d}{g_d}x_i^{ - {\alpha _d}}} }}{{{I_d} + \sigma _d^2}},
	\end{align}
	where $ \sum\limits_{i \in \varphi _c} {{P_d}{g_d}x_i^{ - {\alpha _d}}} $ represents the sum of received signal power from the DBSs in $ \varphi_c $. We use $ \sigma_d^2 $ to denote the power of the additive white Gaussian noise (AWGN), and no specific distribution is assumed in general. $ I_d $ is the aggregate interference from the DBSs outside of $ \varphi_c $ and $ I_d $ can be expressed as 
	\begin{align} \label{interfer}
	{I_d} = \sum\limits_{i \in \left\{ {\varphi _{_D}^t\backslash {\varphi _c}} \right\}} {{P_d}{g_d}x_i^{ - {\alpha _d}}}. 
	\end{align}
	%
	\subsection{Performance Metrics}
	
	In order to enable advanced C-V2X applications such as automated driving applications, stream media \cite{jiao2022spectral}, it is crucial to ensure that the downlink transmission is both reliable and capable of transmitting data at a high rate. This is important not only from the perspective of a single vehicle but also from the perspective of the whole C-V2X network. To this end, this paper conducts an analytical evaluation for coverage probability as follows.
	\begin{itemize}
		\item The coverage probability of the typical vehicle $ v_o $ in downlink, is defined as the probability that the received SINR outperforms a predetermined threshold $ t $. It can be expressed as 
		\begin{align}
		{\mathbb{P}_{{\mathop{ cov}} }}\left( t \right) = \mathbb{P}\left( {S\!I\!N\!{R}_D >t } \right).
		\end{align}
		It can also be calculated as the proportion of vehicles that the received $ S\!I\!N\!R_D $ outperforms a threshold $ t $, i.e., establish a successful connection between the DBS in $ \varphi _c $ and the typical vehicle, among all vehicles in the simulation scenario. Because the cumulative distribution function (CDF) of $ S\!I\!N\!R_D $ is 	$ {\mathbb{P}_{{\mathop{ cov}} }}\left( t \right) = \mathbb{P}\left( { S\!I\!N\!R_D  <t } \right)  $, the coverage probability can also be expressed as the complementary cumulative distribution function (CCDF) of the $ S\!I\!N\!R_D $ at the typical vehicle from the DBSs.

	\end{itemize}
	
	\section{Performance Analysis}
	
	
	We first derive the expression for the joint distance distribution from $ x_1 $ to $ x_n $ in this section. Subsequently, we utilize the results obtained from previous sections to derive the coverage probability and spectral efficiency of C-V2X in a multi-connectivity scenario.
	
	\subsection{The joint distance distribution of the typical vehicle to its service DBS set}
	
	Since the typical vehicle is connected to the $ n $ nearest DBSs in multi-connectivity, no other DBSs are closer than distance $ x_n $. And it also means that all interference DBSs are farther than $ x_n $. The above definition can be expressed by $ f\left( {{x_1},{x_2}, \cdots, {x_n}} \right) $, and we call it joint distance distribution for $ {{x_1},{x_2}, \cdots, {x_n}} $.
	\begin{lemma} \label{joint distance distribution}
		The joint distance distribution of the typical vehicle to its service DBS set $ \varphi{_c} $  from $ x_1 $ to $ x_n $ is 
		\begin{align} \label{f1n}
		f\left( {{x_1},{x_2}, \cdots, {x_n}} \right) = {\left( {2{\lambda _D}} \right)^n}{e^{ - 2{\lambda _D}{x_n}}},
		\end{align}
		where $ x_n $ denotes the distance between the typical vehicle and the $ n $-th closest DBS in $ \varphi_c $.
	\end{lemma}
	\begin{IEEEproof}
		The null probability of a PPP in an area $ A $ is $e^{-\lambda A}$, where $A=2\lambda x$ in 1-D PPP and $ A= \pi x^2 $ in 2-D PPP, thus the complementary cumulative distribution function (CCDF) of $ x_1 $ is 
		\begin{align}
		\mathbb{P}\left [ x >x_1 \right ] &= \mathbb{P}\left [ \text{no~DBS~closer~than~} x_1 \right ] \nonumber\\
		&=e^{-2\lambda_D x_1}.
		\end{align}
		Because the cumulative distribution function (CDF) $=1- $CCDF, the CDF of $ x_1 $ is
		\begin{align}
		F\left( x_1 \right) = 1 - {e^{ - 2{\lambda _D}x_1}}.
		\end{align} 
		Since the probability density function (PDF) $f\left ( x  \right ) =\frac{\partial F\left (x  \right )}{\partial x} $ \cite{chiu2013stochastic}, the PDF of $ x_1 $ is
		\begin{align}
		f\left( x_1 \right) = 2{\lambda _D}{e^{ - 2{\lambda _D}x_1}}.
		\end{align}
		
		According to the definition of Section 3.3 in \cite{moltchanov2012distance},
		let $ f\left( {{x_2}|{x_1}} \right) $ denote the probability that the 2nd closest DBS is at $ x_2 $ given that the closest one is at the distance of $ x_1 $. Thus the probability of having no DBSs between the distances $ x_1 $ and $ x_2 $ can be calculated as follows
		\begin{align} \label{f1}
		f\left( {{x_2}|{x_1}} \right) = 2{\lambda _D}{e^{ - 2{\lambda _D}\left( {{x_2} - {x_1}} \right)}}.	\end{align}
		
		According to the conditional probability Bayes theorem\cite{berrar2018bayes}, the $ f\left( {{x_2},{x_1}} \right) $ denotes the joint distance distribution to two nearest distances, i.e., the probability of having at least one point in ($ x_2+\triangle x $, $ \triangle x $ is an infinitesimal quantity) is 
		\begin{align}\label{f12}
		f\left( {{x_1},{x_2}} \right) = f\left( {{x_2}|{x_1}} \right)f\left( {{x_1}} \right) = {\left( {2{\lambda _D}} \right)^2}{e^{ - 2{\lambda _D}{x_2}}}.\end{align}
		By following the similar procedures in Eq. \eqref{f1} and Eq. \eqref{f12}, the joint distance distribution $ f\left( {{x_1},{x_2} \cdots {x_n}} \right) $ from $ x_1 $ to $ x_n $ is
		\begin{align} \label{jpdf}
		f\left( {{x_1},{x_2}, \cdots, {x_n}} \right) = {\left( {2{\lambda _D}} \right)^n}{e^{ - 2{\lambda _D}{x_n}}}.
		\end{align}
	\end{IEEEproof}
	
	PDF is an important performance indicator for communication networks. To compare the joint distance distribution  $f\left( {{x_1},{x_2}, \cdots, {x_n}} \right) $  and the PDF of $ x_n $, we provide the PDF of $ x_n $ in Eq. \eqref{pdfxn} as
	\begin{align} \label{pdfxn}
	f\left( {{x_n}} \right) = \frac{{{{\left( {2{\lambda _b}{x_n}} \right)}^n}}}{{{x_n}\Gamma \left( n \right)}}{e^{ - 2{\lambda _b}{x_n}}}, \end{align} 
	where when the argument of the Gamma function is a positive integer, the value of the gamma function can be expressed as the factorial of that argument minus 1, $ \Gamma ( n )= (n-1)! $.

	\subsection{Coverage Probability}
	
	
	A general expression for the coverage probability of multi-connectivity in C-V2X is calculated in this subsection.
	\begin{theorem} \label{CP}
		A vehicle is considered to be within the coverage area if its $ S\!I\!N\!R_D $ value from the nearest base station exceeds a certain threshold value $ t $. On the other hand, if the $ S\!I\!N\!R_D $ falls below $ t $, the vehicle is dropped from the network.
		Thus, the coverage probability of downlink for multi-connectivity C-V2X is
		\begin{align}
		&\mathbb{P}\left( {{{S\!I\!N\!}}{{{R}}_D} > {{t}}} \right)\nonumber\\
		&= \int_{0 < {x_1} < {x_2} <  \cdots  < {x_m} < \infty } {{\zeta _{I_D^{}}}\left( j \right)\exp \left( { - \frac{{\mu t\sigma _d^{{2}}}}{{\sum\limits_{{{i = 1}}}^{{m}} {{P_d}x_i^{{{ - }}{\alpha _d}}} }}} \right) \times } \nonumber\\
		&f\left( {{x_1},{x_2}, \cdots ,{x_m}} \right)d{x_1}d{x_2} \cdots d{x_m},
		\end{align}
		where $ j = \frac{\mu t}{{\sum\limits_{{{i = 1}}}^{{m}} {{P_d}x_i^{{{ - }}{\alpha _d}}} }} $, $ m $ is the number of cooperating DBSs in the cooperative set, $ \zeta _{I_D^{}}\left( j \right)  $ is the Laplace transform of random variable interference $ I_D $ evaluated at $ j $ and $ \zeta _{I_D^{}}\left( j \right)  $ is
		\begin{align} 
		&{\zeta _{{I_D}}}\left( j \right)
		\mathop =  \limits^{} \exp \left[ { - 2{\lambda _D}\int_{x_m}^\infty  {1 - \frac{\mu}{{j{P_d}x_i^{ - {\alpha _d}} + \mu}}d{x_i}} } \right].
		\end{align}
	\end{theorem}

	\begin{IEEEproof}
		The proof of coverage probability in downlink is 
		\begin{align}
		&{\mathbb{P}}\left( {{{S\!I\!N\!}}{{{R}}_D} > {{t}}} \right)\nonumber\\
		&{{ \mathop  = \limits^{\left( a \right)} \mathbb{P}}}\left( {\frac{{\sum\limits_{{\rm{i = 1}}}^{\rm{m}} {{P_d}{g_d}x_i^{{\rm{ - }}{\alpha _d}}} }}{{{I_D}{\rm{ + }}\sigma _d^{\rm{2}}}} > {{t}}} \right)\nonumber\\
		&= \mathbb{P}\left( {{g_d} > \frac{{t\left( {{I_D}{\rm{ + }}\sigma _d^{\rm{2}}} \right)}}{{\sum\limits_{{\rm{i = 1}}}^m {{P_D}x_i^{{\rm{ - }}{\alpha _d}}} }}} \right)\nonumber\\
		&\mathop  = \limits^{\left( b \right)} {\mathbb{E}_{{x_i},{I_d}}}\left[ {\exp \left( { - \frac{{\mu t\left( {{I_D}{\rm{ + }}\sigma _d^{\rm{2}}} \right)}}{{\sum\limits_{{\rm{i = 1}}}^{\rm{m}} {{P_d}x_i^{{\rm{ - }}{\alpha _d}}} }}} \right)} \right]\nonumber\\
		&\mathop  = \limits^{\left( c \right)} {\mathbb{E}_{{x_i}}}\left[ {\exp \left( { - \frac{{\mu t\sigma _d^{\rm{2}}}}{{\sum\limits_{{\rm{i = 1}}}^{\rm{m}} {{P_d}x_i^{{\rm{ - }}{\alpha _d}}} }}} \right){\zeta _{I_D^{}}}\left( j \right)} \right]\nonumber\\
		&= \int_{0 < {x_1} < {x_2} <  \cdots  < {x_m} < \infty } {{\zeta _{I_D^{}}}\left( j \right)\exp \left( { - \frac{{\mu t\sigma _d^{\rm{2}}}}{{\sum\limits_{{\rm{i = 1}}}^{\rm{m}} {{P_d}x_i^{{\rm{ - }}{\alpha _d}}} }}} \right) \times } \nonumber\\
		&f\left( {{x_1},{x_2}, \cdots ,{x_m}} \right)d{x_1}d{x_2} \cdots d{x_m},
		\end{align}
		where (a) can obtain the exact expression of $ S\!I\!N\!R_D $ in Eq. \eqref{SINR}. Channel gain $ g_d $ follows an exponential distribution with mean $1/\mu$, thus $ f\left ( g \right ) =\mu  e^{-\mu g} $ in (b). $ \zeta _{I_D^{}}\left( j \right)  $ is the Laplace transform of interference $ I_D $ in (c), and $ j = \frac{\mu t}{{\sum\limits_{{{i = 1}}}^{{m}} {{P_d}x_i^{{{ - }}{\alpha _d}}} }} $. $ f\left( {{x_1},{x_2}, \cdots ,{x_m}} \right) $ is the joint distance distribution in Eq. \eqref{f1n}. 
		Based on the definition of the Laplace transform, the derivation of $ \zeta _{I_D^{}}\left( j \right)$ is
		\begin{align} \label{laplacet}
		&{\zeta _{{I_D}}}\left( j \right) = {\mathbb{E}_{{I_D}}}\left[ {{e^{ - j{I_D}}}} \right]\nonumber\\
		&\mathop  = \limits^{(a)} {\mathbb{E}_{{I_D}}}\left[ {\exp \left( { - j\sum\limits_{i \in \varphi _D^t\backslash \left\{ {{x_1},{x_2}, \cdots ,{x_m}} \right\}}^{} {{P_d}{g_d}x_i^{ - {\alpha _d}}} } \right)} \right]\nonumber\\
		&\mathop  = \limits^{\left( b\right)} {\mathbb{E}_{\Theta _I^d,\left\{ {{g_d}} \right\}}}\left[ {\prod\limits_{i \in \Theta _I^d} {{e^{ - j{P_d}{g_d}x_i^{ - {\alpha _d}}}}} } \right]\nonumber\\
		&\mathop  = \limits^{\left( c \right)} \exp \left[ { - 2{\lambda _D}\int_{{x_m}}^\infty  {1 - } } \right.\nonumber\\
		&\left. {{\mathbb{E}_{{g_d}}}\left[ {\exp \left( { - j{P_d}{g_d}x_i^{ - {\alpha _d}}} \right)} \right]d{x_i}} \right]\nonumber\\
		&\mathop  = \limits^{(d)} \exp \left[ { - 2{\lambda _D}\int_{x_m}^\infty  {1 - \frac{\mu}{{j{P_d}x_i^{ - {\alpha _d}} + \mu}}d{x_i}} } \right],
		\end{align}
		where we use $ {\Theta _I^d}= \varphi _D^t\backslash \left\{ {{x_1},{x_2}, \cdots ,{x_m}} \right\} $ to denote the interference DBSs, interference $ I_D $ can be obtained in Eq. \eqref{interfer}. (b) follows that the accumulative term on a subject to the exponent can be changed into a cumulative form. (c) is derived from the probability generating functional (PGFL) of the PPP, i.e., 
		\begin{align}
		\mathbb{E}\left ( \prod f\left ( x \right )  \right ) =exp\left ( -\lambda \int_{R^2}\left ( 1-f\left ( x \right )  \right ) dx  \right ).
		\end{align}  
		$ \mathbb{E}_{{g_d}}\left[ {\exp \left( { - j{P_d}{g_d}x_i^{ - {\alpha _d}}} \right)} \right] $ follows the PDF of $ g_d $ in (d) and the specific derivation is
		\begin{align}
		&{\mathbb{E}_{{g_d}}}\left[ {\exp \left( { - j{P_d}{g_d}x_i^{ - {\alpha _d}}} \right)} \right]\nonumber\\
		&= \int_0^\infty  {{e^{ - j{P_d}{g_d}x_i^{ - {\alpha _d}}}}{\mu e^{ - \mu{g_d}}}} d{g_d}\nonumber\\
		&=  - \frac{{{\mu e^{ - j\left( {{P_d}x_i^{ - {\alpha _d}} + \mu} \right){g_d}}}}}{{j{P_d}x_i^{ - {\alpha _d}} + \mu}}\left| {_0^\infty } \right.\nonumber\\
		&= \frac{\mu}{{j{P_d}x_i^{ - {\alpha _d}} + \mu}}
		\end{align}
		Since the farthest cooperation DBS is at a distance of $ x_m $, the integration limits are from $ x_m $ to $ \infty  $ in (d).

	\end{IEEEproof}

		\begin{figure}[t]
		\centering
		\centerline{\includegraphics[width=0.77\hsize]{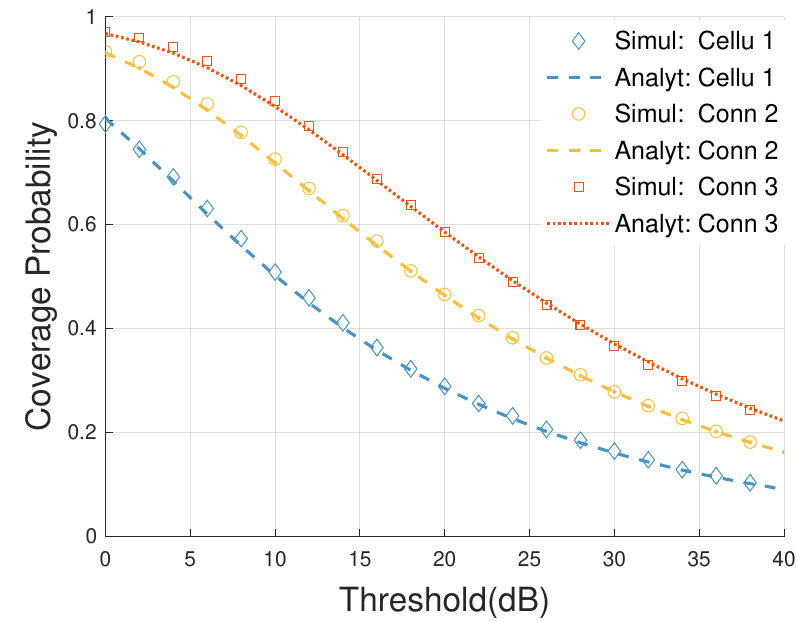}}
		\caption{Coverage probability variation with threshold $ t $.}
		\label{CPT}
	\end{figure}
		\begin{table}[b]
		\centering
		\caption{SYSTEM PARAMETERS}
		\begin{tabular}{l|l }
			\toprule
			\hline
			\label{table1}
			Parameters  & Value \\
			\hline
			The length of road (km) & 300\\
			DBS transmitting power $P_{d}$ (dBm)& 23 \\
			Vehicle transmitting power $P_{V}$ (dBm)& 20 \\		
			Density of vehicle on road $ \lambda_{v}$  (nodes/km)   &20  \\
			Density of DBS $ \lambda_{d}$  (nodes/km)   &5  \\
			Pathloss exponent for downlink  $\alpha _{d}$ &4 \\
			Noise power $\sigma _d^2 $ (dBm)    &-96 \\
			Mean of log-normal shadowing gain (dB) &0 \\
			Std of shadowing gain for MBS (dB) & 2\\
			\hline
			\bottomrule
		\end{tabular}
	\end{table}
	
	\section{NUMERICAL and Simulation RESULTS}
	A communication scenario on a straight urban freeway is considered in this section. The length of the freeway is set as 300 km. We first verify the proposed theoretical derivation in previous sections over 10,000 Monte Carlo simulations of the DBSs and vehicles following 1-D PPPs. We use `Cellu 1', `Conn 2', and `Conn 3' to abbreviate single-connectivity, dual-connectivity, and triple-connectivity, respectively, in the legends of the figures. According to \cite{jiao2022spectral, wu2020performance}, Table \ref{table1} summarizes the system simulation parameters employed in this paper.




	The coverage probability variation of downlink with threshold $ t $ is illustrated in Fig. \ref{CPT}.
	The points represent simulation values and the dashed line represents theoretical values in Fig. \ref{CPT}. It is apparent that the simulation values closely match the theoretical values, which further confirms the validity of the theoretical derivation results.
	The density of BS $ \lambda_{c} $ in single-connectivity is set as 10 nodes/km, and the density $\lambda_d$ of DBS in multi-connectivity is set as 5 nodes/km. Though $\lambda_c > \lambda_{d}$, we can see that the dual-connectivity and triple-connectivity still have a greater coverage probability than single-connectivity. This suggests that multi-connectivity performs better than cellular single-connectivity in C-V2X and multi-connectivity enhances the coverage area of communications.

\begin{figure}[t!]
		\centering
		\centerline{\includegraphics[width=0.75\hsize]{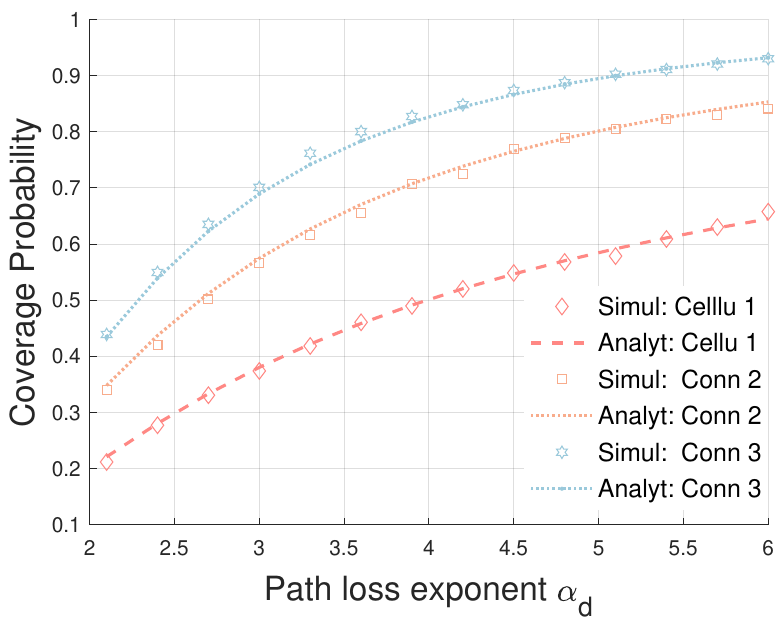}}
		\caption{The effect of path loss exponent on coverage probability.}
		\label{CPA}
	\end{figure}

	Fig. \ref{CPA} illustrates the coverage probability as a function of path loss exponent $\alpha_{d}$. It can be seen that the Monte Carlo simulation data and analytical data fit well. As the path loss exponent increases, the coverage probability increases, primarily due to the fact that the attenuation of received signal power from DBSs $\in \varphi_c$ decreases relative to the interference signal power $ I_D $. This results in an improvement in the SINR, leading to an increase in coverage probability.

	\begin{figure}[t!]
			\centering	
		\includegraphics[width=0.8\hsize]{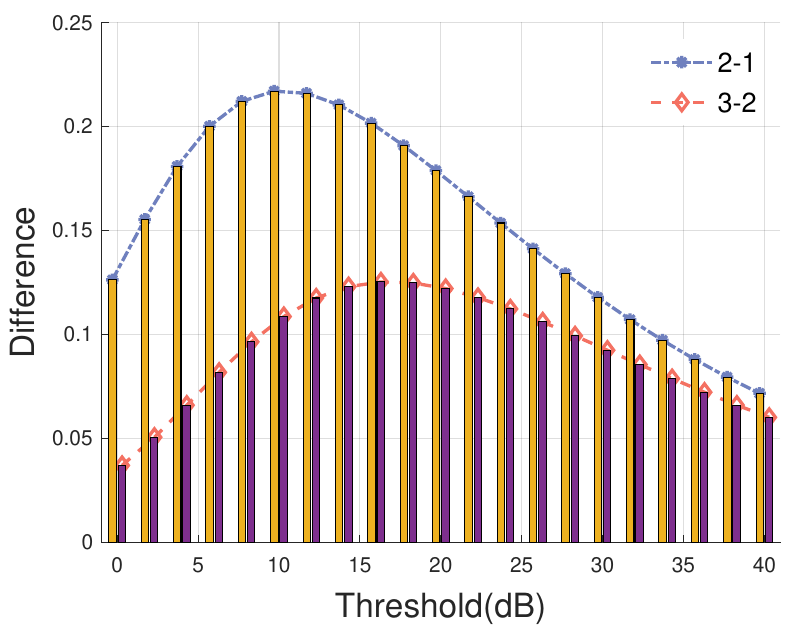}  	 
		\caption{Coverage probability differences variation with threshold $ t$.}
		\label{cp diff}
	\end{figure}

\begin{figure}[!t]
	\centering	
	\includegraphics[width=0.8\hsize]{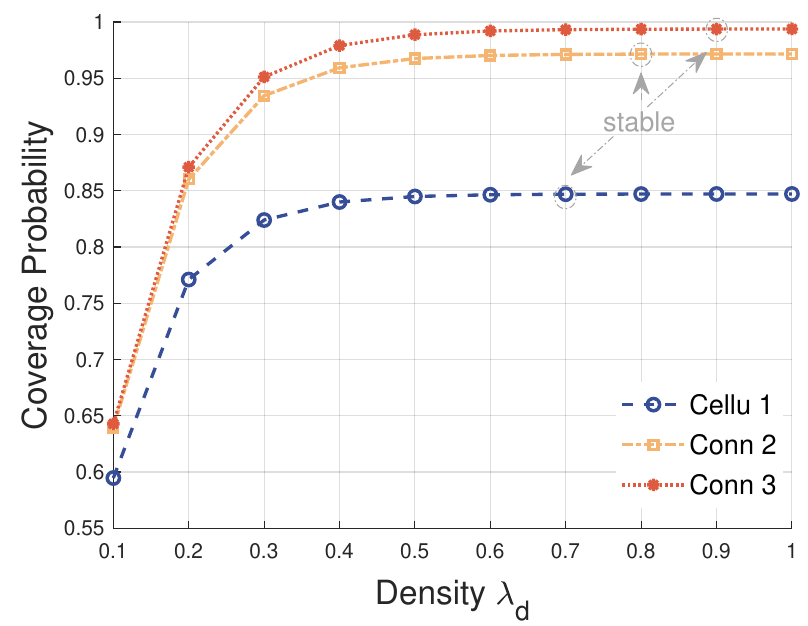}  	 
	\caption{Coverage probability variation with base station density $\lambda_d$.}
	\label{cpdensity}
\end{figure}
	The coverage probability of all cases decreases as the threshold $ t $ increases, while the difference between single-connectivity and multi-connectivity first increases and then decreases in Fig. \ref{cp diff}.
	This is mainly because the coverage probability is respectively very high and relatively low at small and large thresholds. Only when the threshold value is in the middle range, the difference in coverage probability is large, and the advantage of applying multi-connectivity is also demonstrated.
	It can be observed that increasing the number of connected DBSs does not result in a proportional increase in the coverage probability gain.
	Hence, at a particular threshold, there exists a balance between the number of corporation DBSs that are connected and the associated cost.

	As shown in Fig. \ref{cpdensity}, the coverage probability varies with the density $\lambda_d $ of DBS. The coverage probability goes up first and then stays relatively constant for both cellular single-connectivity and multi-connectivity when the density $ \lambda_d $ increases. It can also be seen that the growth rate of multi-connectivity is faster than that of single-connectivity, and it also reaches a stable point later. At the same time, it can be seen that the coverage probability does not increase significantly with the addition of tripe-connectivity compared to dual-connectivity. Therefore, increasing the number of cooperative BSs can improve the communication coverage area, but it may also incur high costs.

	\section{CONCLUSION}
	
	This paper proposed a multi-connectivity performance analysis framework to enhance network performance in C-V2X.
	 Through analyzing the coverage probability in the downlink, this paper has provided insights into the effect of path loss exponent and DBS density on coverage probability. The extensive Monte Carlo simulations have effectively validated the proposed framework and demonstrated the effectiveness of multi-connectivity technology in enhancing the performance of C-V2X networks. The findings of this paper have important implications for the research and practical applications of multi-connectivity C-V2X in the 5G era, and further investigations are warranted to explore the full potential of this technology for next-generation communication systems.

	\section*{Acknowledgment}
	This work is supported in part by the  National Natural Science Foundation Original Exploration Project of China under Grant 62250004, the National Natural Science Foundation of China under Grant 62271244, the Natural
	Science Fund for Distinguished Young Scholars of Jiangsu Province under Grant BK20220067.

	\balance
	\bibliographystyle{IEEEtran}
	
	\bibliography{references}

\end{document}